\documentclass[manuscript]{acmart}
\AtBeginDocument{%
  \providecommand\BibTeX{{%
    \normalfont B\kern-0.5em{\scshape i\kern-0.25em b}\kern-0.8em\TeX}}}
\usepackage{graphicx,tabularx,subcaption}
\usepackage{booktabs}
\usepackage{multirow}
\usepackage[capitalise,nameinlink,compress]{cleveref}
\usepackage{xcolor}
\usepackage{comment}
\usepackage[english]{babel}
\usepackage{fontawesome5}
\usepackage{arydshln}
\usepackage{tikz}
\usepackage{enumitem}
\graphicspath{{images/}}
\usepackage{subcaption}

\definecolor{main}{HTML}{aa72d4}    
\definecolor{sub}{HTML}{f2e8f8}     

\definecolor{greenShade}{RGB}{239, 255, 232}
\definecolor{greenFont}{RGB}{34, 139, 34}

\definecolor{blueShade}{RGB}{231, 239, 253}
\definecolor{blueFont}{RGB}{19, 99, 223}

\definecolor{purpleShade}{RGB}{241, 232, 248}
\definecolor{purpleFont}{RGB}{112, 48, 160}

\definecolor{yellowShade}{RGB}{255, 253, 235}
\definecolor{yellowFont}{RGB}{204, 160, 29}

\definecolor{orangeShade}{RGB}{252, 230, 208}
\definecolor{orangeFont}{RGB}{249, 111, 7}

\DeclareRobustCommand\colorcircle[2]{\tikz[baseline=(char.base)]{\node[shape=circle,fill={#1},inner sep=0pt, minimum size=1em, font=\footnotesize] (char) {\textcolor{white}{#2}}}}

\copyrightyear{2024} 
\acmYear{2024} 
\setcopyright{acmlicensed}\acmConference[UMAP Adjunct '24]{Adjunct Proceedings of the 32nd ACM Conference on User Modeling, Adaptation and Personalization}{July 1--4, 2024}{Cagliari, Italy}
\acmBooktitle{Adjunct Proceedings of the 32nd ACM Conference on User Modeling, Adaptation and Personalization (UMAP Adjunct '24), July 1--4, 2024, Cagliari, Italy}
\acmDOI{10.1145/3631700.3664910}
\acmISBN{979-8-4007-0466-6/24/07}

\begin{document}


\title[Representation Debiasing of Generated Data]{Representation Debiasing of Generated Data Involving Domain Experts}

\author{Aditya Bhattacharya}
\orcid{0000-0003-2740-039X}
\email{aditya.bhattacharya@kuleuven.be}
\affiliation{%
  \institution{KU Leuven}
  \city{Leuven}
  \country{Belgium}
}

\author{Simone Stumpf}
\orcid{0000-0001-6482-1973}
\email{Simone.Stumpf@glasgow.ac.uk}
\affiliation{%
  \institution{University of Glasgow}
  \city{Glasgow}
  \country{Scotland, UK}
}

\author{Katrien Verbert}
\orcid{0000-0001-6699-7710}
\email{katrien.verbert@kuleuven.be}
\affiliation{%
  \institution{KU Leuven}
  \city{Leuven}
  \country{Belgium}
}

\renewcommand{\shortauthors}{Bhattacharya, et al.}

\begin{abstract}
Biases in Artificial Intelligence (AI) or Machine Learning (ML) systems due to skewed datasets problematise the application of prediction models in practice. Representation bias is a prevalent form of bias found in the majority of datasets. This bias arises when training data inadequately represents certain segments of the data space, resulting in poor generalisation of prediction models. Despite AI practitioners employing various methods to mitigate representation bias, their effectiveness is often limited due to a lack of thorough domain knowledge. To address this limitation, this paper introduces human-in-the-loop interaction approaches for representation debiasing of generated data involving domain experts. Our work advocates for a controlled data generation process involving domain experts to effectively mitigate the effects of representation bias. We argue that domain experts can leverage their expertise to assess how representation bias affects prediction models. Moreover, our interaction approaches can facilitate domain experts in steering data augmentation algorithms to produce debiased augmented data and validate or refine the generated samples to reduce representation bias. We also discuss how these approaches can be leveraged for designing and developing user-centred AI systems to mitigate the impact of representation bias through effective collaboration between domain experts and AI. 
\end{abstract}




\begin{teaserfigure}
  \centering
  \includegraphics[width=0.65\linewidth]{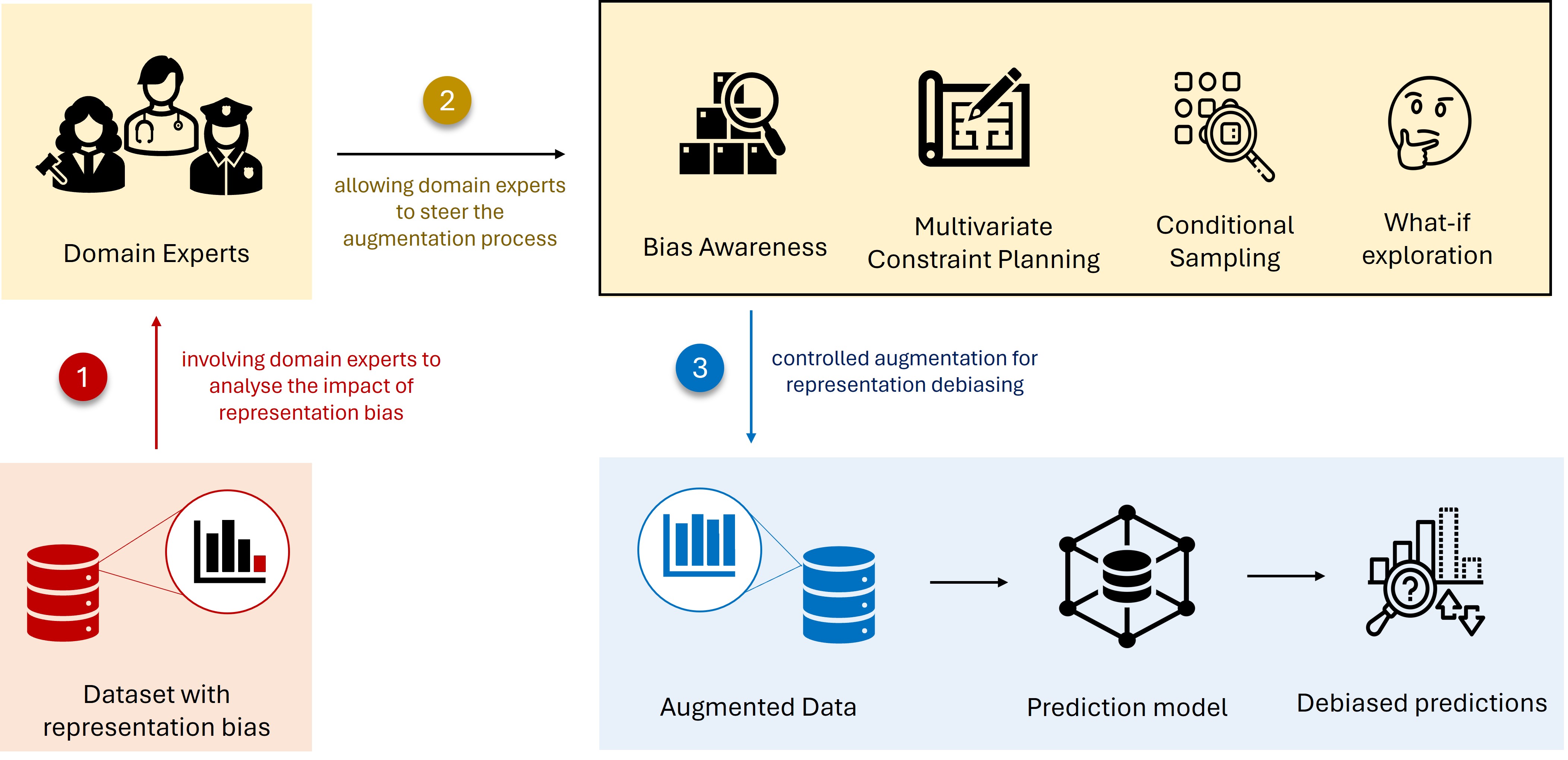}
  \caption{This paper presents the interaction approaches for representation debiasing of generated data, highlighting the crucial role of domain experts in steering data augmentation methods to mitigate representation bias within datasets. }
  \Description[Representation debiasing of generated data]{This paper presents our proposed interaction approaches for generated data, highlighting the crucial role of domain experts in steering data augmentation methods to mitigate representation bias within datasets.}
  \label{fig:teaser_image}
\end{teaserfigure}

\maketitle

\section{Introduction}
The impact of Artificial Intelligence (AI) has been significant across nearly every application domain. However, the quality of the AI models largely depends on the quality of the datasets used to train them \cite{2019DataQA, Ding2018, Shahbazi2023, aldoseri2023rethinking, mehrabi2022survey}. Moreover, several past incidents highlight the devastating consequences of using biased and erroneous datasets for training AI models, such as discriminatory treatment of users based on demographic characteristics like gender, age, race, and religion by AI systems \cite{2019DataQA, khan2023drawbacks, ahmad2023impact, BhattacharyaXAI2022, armstrong2014errors, darksideAI2023}. Therefore, this increasing emphasis on data has led to a shift towards data-centric AI \cite{Mazumdar2022, zha2023datacentric, jakubik2024datacentric, SINGH2023144} approaches, where researchers and practitioners prioritise mitigating biases and issues in datasets over optimising model architectures.

Within the spectrum of biases encountered in machine learning (ML) systems \cite{mehrabi2022survey}, \textit{representation bias} stands out as pervasive across datasets lacking a systematic data collection process \cite{Shahbazi2023}. Representation bias is defined as the bias in the training data that occurs due to inequality in the proportion of information leading to models that are biased in favour of the majority samples \cite{Shahbazi2023, mehrabi2022survey}. For instance, a dataset captured through a survey to measure screen time on electronic gadgets for teenagers could have representation bias if only high school students are involved and if home-schooled students or dropouts are not involved in the survey. Consequently, if the dataset is not representative of an entire population, the outcome of prediction models used in decision systems may not be trustworthy for the sub-populations that are under-represented. This is because the model has fewer data points to learn from for such under-represented populations, and hence, it does not generalise well for such sub-populations. 

The most effective practical solution to mitigate representation bias is through \textit{data augmentation} \cite{Iosifidis2018DealingWB, Shahbazi2023, sharma2020data}. Data augmentation is a technique used to increase the size and diversity of training datasets by generating new samples while ensuring that the statistical properties of the original dataset are preserved \cite{Shahbazi2023, dataaugment2023}. For instance, if a particular group is underrepresented in the original dataset, data augmentation can be used to generate more synthetic samples of that group. This method essentially allows the prediction model to learn from more samples to avoid overfitting or underfitting problems \cite{dataaugment2023}. By generating more samples of underrepresented groups, the resulting augmented dataset becomes more balanced and representative of the real-world population, mitigating the possibility of bias \cite{Shahbazi2023}. 

However, an uncontrolled data augmentation, i.e., the process of applying augmentation algorithms without strict constraints or predefined rules, can still retain the representation bias in the data as these methods try to preserve the distribution and the statistical properties of the original dataset \cite{alkhawaldeh2023challenges, bareła2023dataaugment, balestriero2022effects, superb-ai-blog}. Additionally, data augmentation using up-sampling methods like SMOTE \cite{Chawla_2002} has been criticised for increasing data issues like data drift, correlated features, duplicates, or even outliers \cite{limitationsmote2022, alkhawaldeh2023challenges}. 

Therefore, domain knowledge is important for applying constraints to avoid the generation of problematic samples. It is also necessary for identifying, modifying and removing synthetic samples that may add more bias. Thus, domain knowledge can be crucial in the debiasing process of representation bias, highlighting the importance of involving domain experts in the process of representation debiasing in ML systems.

To address the current limitations in mitigating representation bias, this paper introduces our interaction approaches for representation debiasing of generated data by involving domain experts to steer data augmentation methods. These approaches aim to facilitate domain experts in explaining the presence of representation bias in datasets and leverage their expertise to generate unbiased augmented samples. Thus, interaction systems offering these interaction approaches can reduce representation bias in their training datasets and consequently, improve the overall prediction models and the quality of the training datasets for unbiased predictions. Furthermore, through an exploratory user study, we have instantiated these approaches into designs of visual components for the healthcare domain. Therefore, these interaction approaches can take us one step closer towards achieving debiased user-centred AI systems.

\section{Background and Related Work}
\subsection{Representation Bias}
As discussed earlier, representation bias arises due to the absence of sufficient information for the various subgroups or sub-categories in the dataset \cite{Shahbazi2023}. This type of bias can be introduced from a number of factors such as historical trends observed in past events, skewness in the distribution of the data, procedures used for data preparation and acquisition, and selection and sampling biases \cite{Shahbazi2023, mehrabi2022survey}. Representation bias is also one of the main concerns for achieving group or sub-group fairness, as models trained on data lacking representation from certain groups or subgroups are expected to exhibit lower accuracy when applied to these under-represented categories \cite{Shahbazi2023}.
Thus, training datasets must include sufficient samples from the ``less popular segments'' of the data space in order for the ML system to handle these segments well. Yet, representation bias is considered to be one of the critical problems in ML systems, leading to biased and unfair predictions \cite{Shahbazi2023}.

Representation bias is measured through two main metrics: (1) Representation rate and (2) Data coverage \cite{Shahbazi2023}. Representation rate of a sub-group is defined as the ratio of its sample counts to the highest frequency of occurrence among all sub-groups. Suppose if variable $A$ has three sub-groups: $x$, $y$ and $z$, such that $x + y + z = N$, where $N$ is the total number of samples for variable $A$. Then, the representation rate of sub-group $x$ is measured by: $r_x = \frac{x}{max(x, y, z)}$. 
For instance, let us consider a variable representing education level of job applicants that has three sub-categories: high-school degree, bachelor's degree and master's degree. Suppose that the dataset has 600 samples such that the number of applicants with education level as high-school degree is 100, bachelor's degree is 300 and master's degree is 200. Then, the representation rate of applicants with a high-school degree is: $r_1 = \frac{100}{max(100, 300, 200)} = 0.33$. Similarly, the representation rates of the other two sub-groups are 1.0 and 0.67, respectively. This implies that applicants with a high school degree are less represented than applicants with bachelor's or master's degrees in the dataset.

Meanwhile, data coverage is defined as the minimum number of samples that should be present for each sub-category. For instance, if the data coverage is set as 150, then the sub-category of applicants with an education level of a high-school degree does not meet the data coverage criteria as it has only 100 samples. 
Regardless of the data space, it is critical to have a high enough coverage for all significant sub-populations in the data to ensure their adequate representation.

\subsection{Data Augmentation}

To address representation bias, AI practitioners have recommended improving the data collection process and gathering additional data for under-represented segments of the dataset \cite{Shahbazi2023, dataaugment2023}. Nonetheless, due to practical limitations and challenges, collecting new data may not always be feasible \cite{dataaugment2023}. Thus, researchers have proposed generating synthetic samples for the under-represented segments as a debiasing approach for representation bias \cite{Shahbazi2023, debiasaugment}. Data augmentation is the process of generating new data points by applying various transformations or modifications to the existing data \cite{dataaugment2023, Shahbazi2023, bareła2023dataaugment}. 
AI practitioners have also relied on data augmentation methods to address the class imbalance problem, which arises when the target variable predominantly contains samples from one of its sub-categories \cite{temraz2021solving}. Methods such as SMOTE \cite{Chawla_2002} and ADASYN \cite{ADASYN2008} have been commonly used in such scenarios as they can identify the majority and the minority categories and generate samples for the minority category for creating a balanced dataset.

Despite the significance of such up-sampling techniques, these methods have been criticised for introducing data quality issues such as data drift, duplicate records and outliers \cite{limitationsmote2022, alkhawaldeh2023challenges, balestriero2022effects}. Additionally, many generative AI algorithms, such as GANs \cite{Gan2024} and VAEs \cite{VAE2014}, have also been explored by practitioners for generating synthetic samples of the data. For tabular datasets, their different variants, such as CTGAN and TVAE \cite{ctgan2019}, have been considered better as they introduce fewer data issues. By creating more samples of underrepresented groups, these methods can produce an augmented dataset that is more balanced and representative of the real-world population, reducing the likelihood of bias in prediction models \cite{SDV2016}. 

However, researchers have also noted that an uncontrolled data augmentation process can impact prediction models by generating practically implausible samples \cite{dataaugment2023, tang2020, temraz2021solving}. For example, let us consider an under-represented diabetes prediction dataset that has factors such as glucose, body mass index (BMI), age and obesity level of a patient. An uncontrolled data augmentation process can generate samples where the obesity level is high, yet the BMI is less than 18. But practically speaking, a patient with a BMI less than 18 is considered to be underweight and not obese. Moreover, uncontrolled data augmentation can retain the existing representation bias in the data in an attempt to preserve the statistical properties of the data \cite{MUMUNI_DAC, balestriero2022effects, superb-ai-blog}. 

Domain knowledge is considered crucial for performing controlled data augmentation for representation debiasing as it involves setting constraints to prevent the generation of problematic samples. Our work argues the importance of involving domain experts in the process of representation debiasing. Establishing an effective collaboration between domain experts and AI systems is essential for representation debiasing of generated data.

\subsection{Model Steering involving Domain Experts}
Researchers working in interactive machine learning (IML) have primarily focused on investigating various approaches for user-in-the-loop model steering to improve prediction models \cite{fails2003, kulesza_explanatory_2010, kulesza_principles_2015, teso_leveraging_2022, teso2019, bhattacharya2024exmos}. Previous studies have also examined diverse approaches through which end-users could guide ML systems and actively participate in training, fine-tuning, and debugging prediction models \cite{fails2003, kulesza_explanatory_2010, kulesza_principles_2015, muralidhar2018incorporating, spinner2019explainer, stumpf2009interacting, Guo2022BuildingTI, honeycutt2020soliciting}.

Recent works have shown the importance of the active involvement of domain experts in model steering as they can easily identify and correct training data issues that are essential for steering models \cite{bhattacharya2024exmos, bhattacharya2023_technical_report, Schramowski2020}. It has also been argued that for identifying various types of biases in the data a thorough domain knowledge possessed by domain experts is extremely important \cite{2019DataQA, feuerriegel2020fair}. For instance, rich domain knowledge possessed by medical experts facilitates them in investigating the health records of patients and identifying biases or erroneous data that can lead to biased models.

Although recent works have designed and developed innovative approaches for enabling domain experts to steer prediction models \cite{bhattacharya2024exmos, teso_leveraging_2022, teso2019, Schramowski2020}, prior research has largely overlooked their involvement in the process of generating data through controlled data augmentation. Specifically, leveraging domain experts' prior knowledge presents significant potential for mitigating the risks associated with uncontrolled data augmentation and addressing representation bias \cite{tang2020}. Thus, our work aims to fill this gap and explore how domain experts can be involved in the process of representation debiasing of generated data.

\begin{figure*}[h]
  \centering
  \begin{minipage}[b][4.5cm][t]{0.499\textwidth}
    \centering
    \includegraphics[width=\linewidth, height=4.5cm]{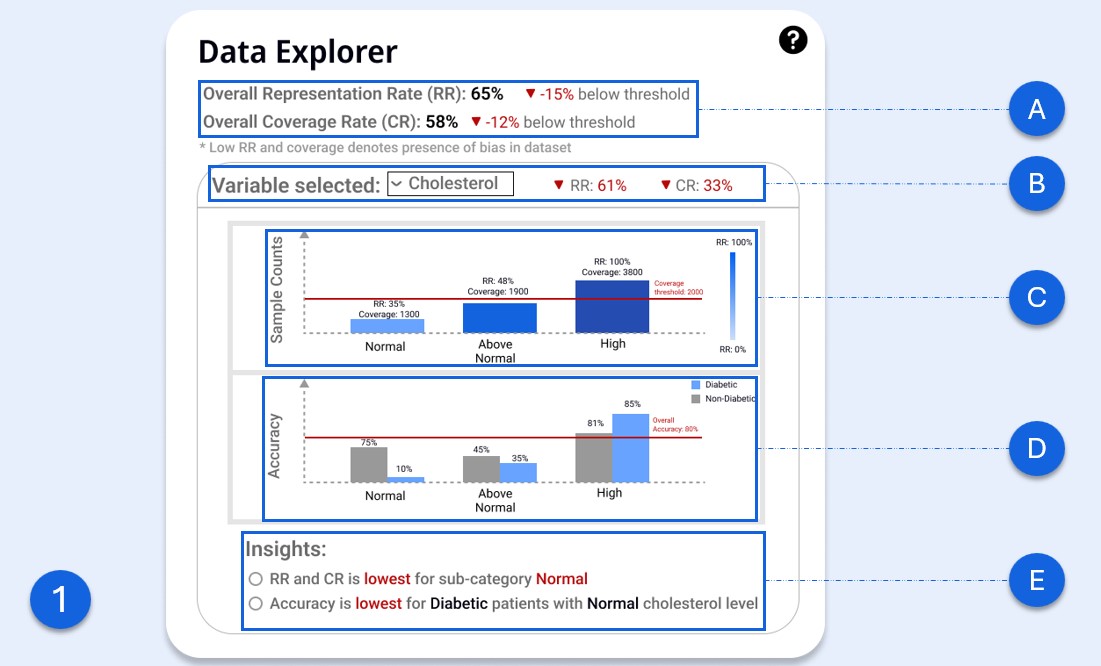}
  \end{minipage}\hfill
  \begin{minipage}[b][4.5cm][t]{0.499\textwidth}
    \centering
    \includegraphics[width=\linewidth, height=4.5cm]{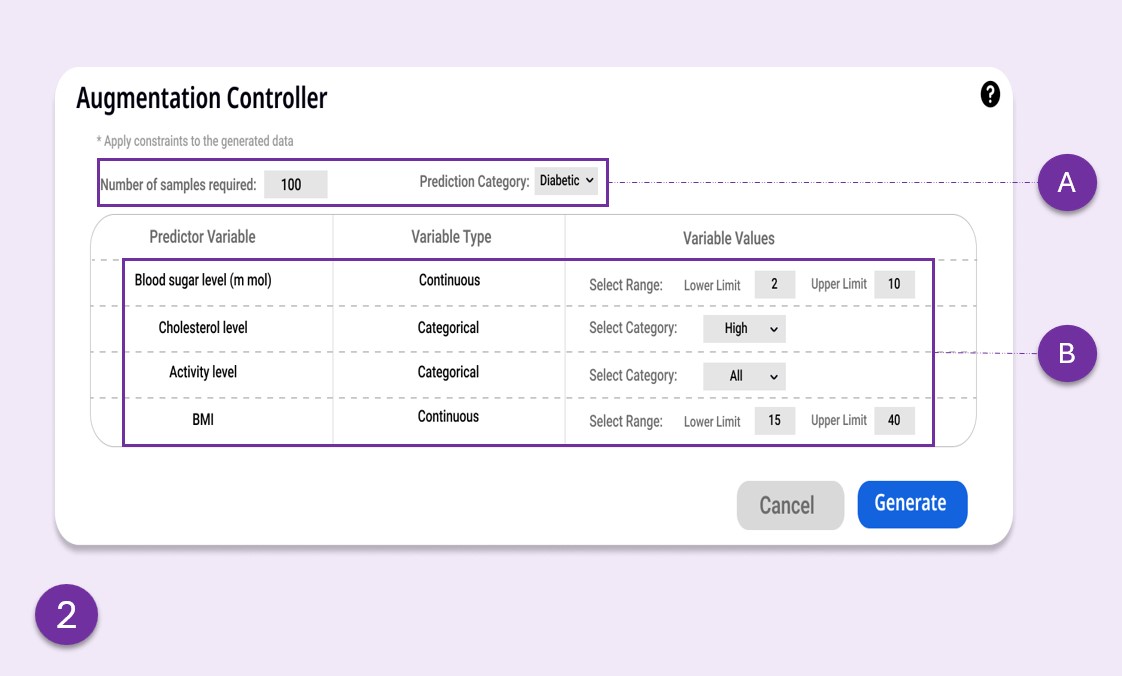}
  \end{minipage}
  
  \begin{minipage}[b][4.5cm][t]{0.499\textwidth}
    \centering
    \includegraphics[width=\linewidth, height=4.5cm]{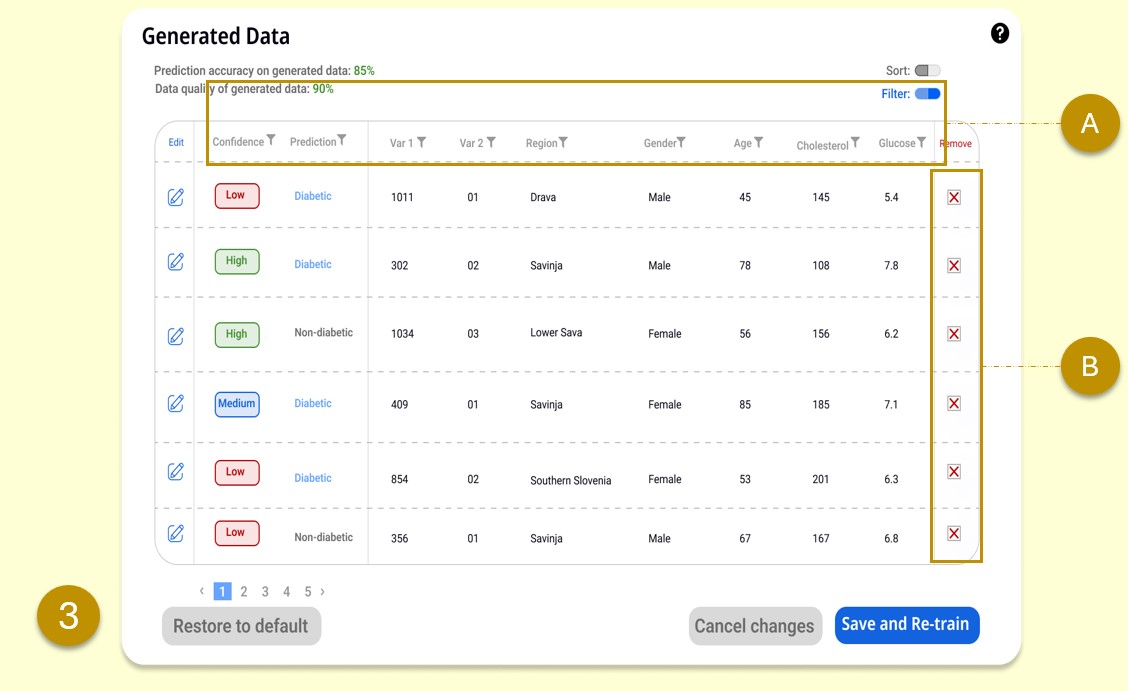}
  \end{minipage}\hfill
  \begin{minipage}[b][4.5cm][t]{0.5\textwidth}
    \centering
    \includegraphics[width=\linewidth, height=4.5cm]{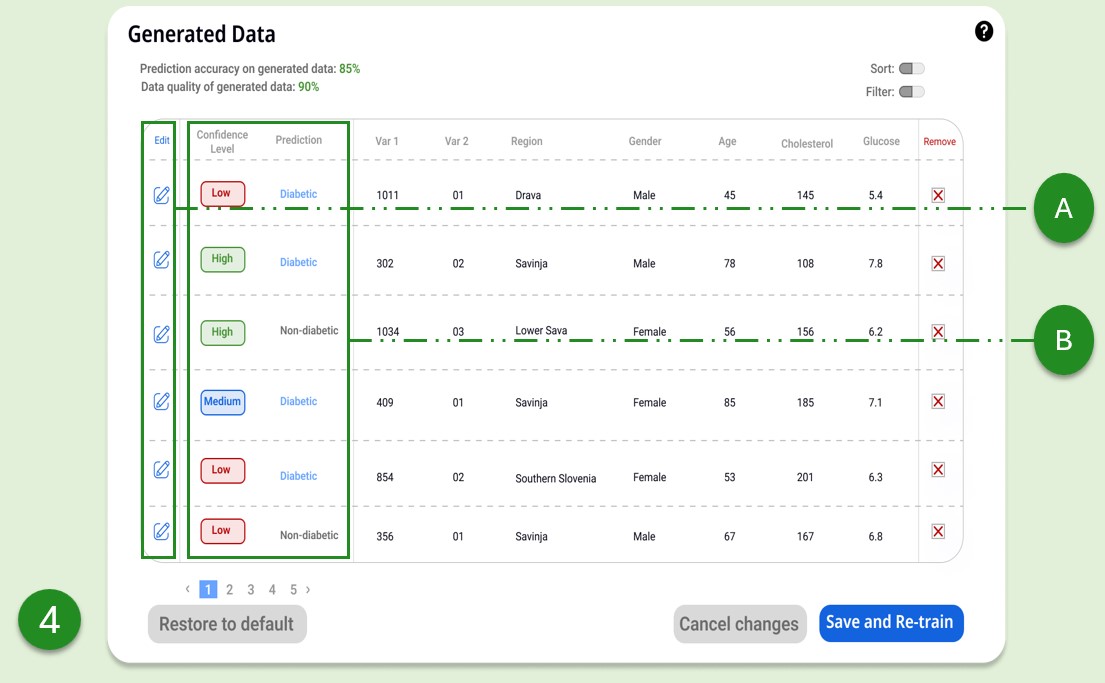}
  \end{minipage}
  \caption{  Design of visual components of a low fidelity prototype instantiating the interaction approaches for representation debiasing. \\  
  \colorcircle{blueFont}{1} \textsc{Bias awareness} involves: \colorcircle{blueFont}{A} presenting the overall representation bias measures, \colorcircle{blueFont}{B} presenting the representation bias measures for each predictor variable, \colorcircle{blueFont}{C} showing the representation bias for each sub-category of the selected predicted variable, \colorcircle{blueFont}{D} showing the corresponding impact on the model performance, and \colorcircle{blueFont}{E} highlighting the most impacted variables and sub-categories.\\
  \colorcircle{purpleFont}{2} \textsc{ Multivariate constraint mapping} involves: \colorcircle{purpleFont}{A} allowing users to specify the required number of generated samples for a specific target class, and \colorcircle{purpleFont}{B} set constraints on the predictor variable values. \\
  \colorcircle{yellowFont}{3} \textsc{Conditional sampling} involves: \colorcircle{yellowFont}{A} allowing users to sample generated data through conditions defined in data filers, and \colorcircle{yellowFont}{B} remove misfit samples.  \\
  \colorcircle{greenFont}{4} \textsc{What-if exploration} involves: \colorcircle{greenFont}{A} allowing users to validate or modify generated samples through ``what-if'' analysis, and \colorcircle{greenFont}{B} applying the prediction model on generated samples to identify problematic samples having low prediction confidence level.}
  \Description[Design of visual components of a low fidelity prototype instantiating the interaction approaches for representation debiasing.]{Design of visual components of a low fidelity prototype instantiating the interaction approaches for representation debiasing: (1) Bias awareness (2) Multivariate constraint mapping (3) Conditional sampling (4) What-if exploration.}
  \label{fig:principles_design}
\end{figure*}

\section{Interaction Approaches for Representation Debiasing}\label{sec_interaction_approaches}
This section presents our interaction approaches for involving domain experts in the representation debiasing of training data used in ML systems. We elaborate on the rationale, description and purpose of these approaches in the following part. We have also conducted an exploratory user study (discussed in \Cref{sec_user_study}) to instantiate these interaction approaches for representation biasing through visual components of a low-fidelity prototype, as illustrated in \Cref{fig:principles_design}. The proposed interaction approaches are as follows:
\begin{itemize}[left=0cm]
    \item \colorbox{blueShade}{ \textcolor{blueFont}{\textsc{Bias awareness:}}} Although domain knowledge is essential to identify representation bias, prior researchers have signified the importance of interactive explanations for assisting domain experts in elucidating the behaviour of prediction systems \cite{Bhattacharya2023, bhattacharya2024exmos, teso_leveraging_2022, lakkaraju2022rethinking}. Thus, we propose an approach for \textit{bias awareness}, that aims at guiding domain experts to identify biased predictor variables with the help of data-centric explanations \cite{anik_data-centric_2021, BhattacharyaXAI2022, Bhattacharya2023}. We recommend allowing domain experts to explore the distributions of categories or sub-categories of predictor variables, including the representation rate and data coverage of each category or sub-category, through interactive visualisations for higher transparency. This interaction approach is aligned with the ML transparency and exploration principles from Bove et al.~\cite{bove_contextualization_2022}, which aims at providing better contextualised explanations for the presence of representation bias. Furthermore, we propose illustrating the model performance corresponding to each category or sub-category of predictor variables to identify those most impacted by representation bias.

    \item \colorbox{purpleShade}{ \textcolor{purpleFont}{\textsc{Multi-variate constraint planning:}}} Despite generating additional samples of underrepresented data, one primary reason for the limited effectiveness of data augmentation algorithms in mitigating representation bias is the generation of practically infeasible samples \cite{dataaugment2023, MUMUNI_DAC}. This occurs because data augmentation algorithms typically treat each predictor variable independently rather than jointly. It requires in-depth domain knowledge to understand the joint impact of the predictor variables. With \textit{multivariate constraint planning}, we propose empowering domain experts to impose constraints on multiple predictor variables. This allows for the generation of specific sets of samples considered essential by experts to mitigate the impact of representation bias. For example, consider the representation debiasing of a diabetes prediction dataset. If healthcare experts identify that only 50 samples of diabetic patients aged 50 to 60 with high cholesterol and high blood pressure are necessary, then multivariate constraint planning can be utilised to achieve their requirement. 
    This interaction approach enables control over the data augmentation process to mitigate the issue of generating practically infeasible data points.
    
    \item \colorbox{yellowShade}{ \textcolor{yellowFont}
    {\textsc{Conditional sampling:}}} The interaction approach of \textit{conditional sampling} is applicable after the application of data augmentation algorithms, allowing domain experts to select only relevant synthetic samples for upsampling the original training data. We recommend providing data filters for setting conditions defined by domain experts, allowing them to identify and remove generated samples which appear to be a misfit. This approach aims at purifying the generated data for minimal introduction of problematic data points during representation debiasing.
    
    \item \colorbox{greenShade}{ \textcolor{greenFont}{\textsc{What-if exploration:}}} The interaction approach for \textit{what-if exploration} of the generated data further allows domain experts to validate the generated samples. It aligns with the concept of ``what-if'' explanations \cite{Lim_CHI_2009, Bhattacharya2023, wang_designing_2019, BhattacharyaXAI2022}, aiming to enhance understanding of the generated samples and their potential impact on prediction models. We recommend applying the prediction model to each generated sample to obtain their predicted target class and the corresponding confidence levels. This can allow domain experts to identify problematic data points that are difficult to train by the prediction algorithm. Additionally, domain experts should be able to adjust the values of generated samples and conduct what-if analysis to rectify such problematic generated instances.
\end{itemize}

\section{Exploratory User Study}\label{sec_user_study}
An exploratory user study was conducted with 5 healthcare experts (2 females, 3 males; age: 29 - 51 years), each having more than four years of experience in healthcare, from \anon{the Faculty of Healthcare Sciences, University of Maribor, Slovenia}, to explore the design space of UI components offering our proposed interaction approaches for representation debiasing. The study included individual co-design and think-aloud sessions (each session lasted for about 30 minutes on average).  The sessions were recorded
and transcribed for analysing the feedback of the participants. The goal of this study was to explore the design space of UI components for the interaction approaches for representation debiasing.

The findings from this exploratory study allowed us to design visual components for a low-fidelity click-through prototype. The design of the visual components is shown in \Cref{fig:principles_design}. In this study, we extensively engaged healthcare experts to gather their perspectives on participating in the task of mitigating representation biases within ML systems. Encouragingly, all participants expressed enthusiastic support for actively contributing to the debiasing process.  Additionally, they have mentioned that this process could allow them to better understand predicted outcomes that do not match their expectations. 
Therefore, as a key takeaway of this study, we came up with the following UI components supporting the four interaction approaches discussed in \Cref{sec_interaction_approaches}:
\begin{enumerate}[start=1,label={\textbf{ \arabic*.}}, left=0cm]
    \item \textsc{Data Explorer} - This component is designed creating \textit{bias awareness} (\Cref{fig:principles_design} \colorcircle{blueFont}{1}). It includes presenting the overall representation bias measures, as well as the representation bias measures for each predictor variable. We included a visual representation of the distribution of each predictor variable, illustrating the representation bias for each sub-category of the selected variable. Additionally, we illustrated the corresponding impact on the model performance for each sub-category and highlighted the most impacted variables and sub-categories.
    \item \textsc{Augmentation Controller} - This component is designed to support \textit{multivariate constraint mapping} (\Cref{fig:principles_design} \colorcircle{purpleFont}{2}). It is designed to allow users to specify the required number of generated samples for a specific target class, and set constraints on the predictor variable values for the generated samples.
    \item \textsc{Generated Data Explorer} - This component is designed to support \textit{conditional sampling} (\Cref{fig:principles_design} \colorcircle{yellowFont}{3}) and \textit{what-if exploration} (\Cref{fig:principles_design} \colorcircle{greenFont}{4}) of generated data. It is designed to allow users to sample generated data through conditions defined in data filers, and remove problematic samples. Moreover, it allows users to validate or modify generated samples through ``what-if'' analysis. By applying the prediction model to generated samples, this component further facilitates users to identify problematic samples having low prediction confidence levels.
\end{enumerate}

\section{Discussions}
\subsection{Debiasing is a continual process}
As noted by prior researchers \cite{Shahbazi2023, superb-ai-blog}, debiasing ML systems is a continual process as new sources of bias can get introduced when the system is deployed in different contexts or deployed with new data (including generated data). 
Although findings from our exploratory study indicate that domain experts can be involved in representation debiasing, their continuous involvement in the debiasing process could be a challenge due to their limited availability. Yet, continuous monitoring and evaluation of prediction models are necessary to ensure that these models remain unbiased and fair. Further investigation is required to analyse how effectively can user-centric AI systems support continuous involvement of domain experts in the debiasing process.

\subsection{Implications on fairness}
The presence of representation bias is one of the primary concerns for group fairness or sub-group fairness of AI systems \cite{Shahbazi2023, mehrabi2022survey, feuerriegel2020fair}. Traditional approaches for mitigating fairness issues involve identifying and removing protected attributes such as gender, race, religion, etc., from the training data for the prediction models \cite{feuerriegel2020fair}. However, only eliminating a protected attribute may not be sufficient if the model has access to other predictor variables that are proxies for the protected attribute \cite{superb-ai-blog}. For instance, if gender is removed as it could be a protected attribute, other attributes such as height, weight or BMI can act as proxy attributes. However, the interaction approaches proposed in this paper can provide a solution to the limitations of these traditional approaches for achieving group or sub-group fairness. Thus, we hypothesise that these interactions could take us closer towards achieving group or sub-group fairness in AI.

\subsection{Future Work}
Future research should consider applying the visual components of representation debiasing to design and develop user-centred AI systems tailored for domain experts. More extensive user studies should be conducted to investigate how these approaches can empower domain experts to improve the overall prediction model performance and data quality of AI systems. Furthermore, it will be interesting to analyse how the process of representation debiasing affects the trust and understanding of domain experts on AI systems.

\section{Conclusion}
To conclude, this paper introduces our proposed interaction approaches for representation debiasing of generated data used in AI systems by allowing domain experts to steer the data augmentation process. In this paper, we describe these approaches, elaborate on the rationale behind them and also present our design of UI components to instantiate these interaction approaches for user-centred AI systems tailored for domain experts. We aim to discuss the implications of these UI components for human-in-the-loop AI systems to mitigate representation bias in the workshop and gain valuable feedback for proposing a framework for representation debiasing, enabling user-centred AI practitioners to take a step closer towards achieving debiased systems.

\begin{acks}
We thank Maxwell Szymanski and Robin De Croon for their valuable feedback on this research. This research was supported by the Flanders AI Research Program and Research Foundation–Flanders (FWO grants G0A4923N and G067721N) and KU Leuven Internal Funds (grant
C14/21/072)~\cite{BhattacharyaCHIDC}. We also thank our participants from the University of Maribor for their feedback captured from the exploratory user study.
\end{acks}

\bibliographystyle{ACM-Reference-Format}
\bibliography{references}


\end{document}